\documentclass[prl,aps,twocolumn,floatfix]{revtex4}
\usepackage{amsmath,graphics,epsfig,color,verbatim}
\usepackage[all,dark]{draftcopy}

\begin{document}

\title{The low-energy ARPES and heat capacity of Na$_{0.3}$CoO$_2$: A DMFT study}

\author{C. A. Marianetti$^1$, K. Haule$^1$, and O. Parcollet$^2$ }

\date{\today}

\begin{abstract}
We use the dynamical mean-field theory (DMFT) to calculate the angle resolved
photoemission spectrum (ARPES) and heat capacity for Na$_{0.3}$CoO$_2$. Both
the traditional Hirsch-Fye Quantum Monte-Carlo technique and the newly
developed continuous time quantum Monte-Carlo technique are used to solve the
DMFT impurity problem. We show that the e$_g$' hole pockets on the Fermi
surface are suppressed  as the on-site coulomb repulsion is increased.  A
quantitative comparison with ARPES experiments and bulk heat capacity
measurements indicate that the on-site coulomb repulsion is large relative to
the LDA bandwidth.

\end{abstract}

\address{$^1$Department of Physics and Astronomy and Center for Condensed Matter
Theory, Rutgers University, Piscataway, NJ 08854--8019}
\address{$^2$ Service de Physique Theorique, CEA/DSM/SPhT-CNRS/SPM/URA 2306
CEA/Saclay, F-91191 Gif-Sur-Yvette, France}

\maketitle

The cobaltates have demonstrated a wide variety complex behavior.  The Na rich
region of the phase diagram displays various degrees of anomalous behavior, such
as Curie-Weiss behavior near a band insulator\cite{Foo:2004}, charge
disproportionation\cite{Mukhamedshin:2005}, and non-Fermi-liquid behavior in the
resistivity\cite{Foo:2004}.  Alternatively, the Na poor region of the phase
diagram appears to be a Fermi-liquid. The magnetic susceptibility
displays Pauli behavior, the resistivity is roughly quadratic at low
temperatures\cite{Foo:2004}, and the system appears to be
homogeneous\cite{Mukhamedshin:2005}. Therefore, the Na poor region of the phase
diagram seems like a natural starting point to attempt to explain the ARPES
experiments and heat capacity measurements from a quantitative standpoint.

In Na$_x$CoO$_2$, the cubic component of the oxygen crystal field splits the Co $d$ manifold into
a set of 3-fold $t_{2g}$ orbitals and 2-fold $e_g$ orbitals, while the trigonal
component will further split the $t_{2g}$ orbitals into $a_{1g}$ and $e_g'$.
The nominal valence of Co in this system will be $4-x$, so the
Fermi-energy will fall within the $t_{2g}$ manifold.  The LDA band structure
displays two degenerate eigenvalues and one non-degenerate eigenvalue at the
$\Gamma$-point, corresponding to the $e_g'$ and $a_{1g}$ eigenvectors.  The
splitting between the eigenvalues is roughly 1 $eV$ with the $e_g'$ levels
below the Fermi energy and the $a_{1g}$ above. Despite this distinct splitting
at the $\Gamma$-point, the on-site orbital energies are nearly degenerate.
Additionally, the projected density-of-states (DOS) clearly show that the
$a_{1g}$ orbital character is strongest at the top and bottom of the band while
the $e_g'$ is present through most of the energy range of the $t_{2g}$  bands.
The Fermi surface consists of a large $a_{1g}$ pocket around the $\Gamma$-point
and six small $e_g'$ satellite pockets\cite{Singh:2000}. 

Several experimental ARPES studies have been performed for
Na$_{0.3}$CoO$_2$
\cite{Yang:2005,Qian:2006,Qian:2006B,Qian:2006C}. 
A general caricature of the LDA bands can be seen in the ARPES. The most notable
difference as compared to LDA is the significant narrowing of the bands, and the
suppression of the $e_g'$ pockets below the Fermi energy. Two previous studies
addressed the effect of correlations on the electronic structure for $x=0.3$,
and they reached completely opposite conclusions.  Zhou et al performed
Gutzwiller calculations for a three-band model corresponding to the LDA $t_{2g}$
band structure\cite{Zhou:2005}. Using an infinite on-site coulomb repulsion,
they show that the quasi-particle bands are significantly narrowed and the
$e_g'$ hole pockets are pushed beneath the Fermi surface.  Although the removal
of the $e_g'$ pockets agrees with the ARPES experiments, it is not clear 
if $U=\infty$ is an excessive assumption and therefore
smaller values of the on-site coulomb repulsion must be considered.  Ishida et
al\cite{Ishida:2005} performed DMFT calculations for the three-band
$t_{2g}$  states of the cobaltates and found that electronic correlations narrow
the bands and enhance the $e_g'$ hole pockets, completely opposite to what was found by
Zhou et al.  

Singh et al. have proposed that the inclusion of the realistic ordering of the
Na destroys the $e_g'$  hole pockets, as demonstrated by LDA calculations for
Na$_{0.7}$CoO$_2$\cite{Singh:2006}.  Given that the pockets will inevitably be
destroyed as $x \rightarrow 1$  because the chemical potential moves towards the
top of the t2g bands, Na$_{0.7}$CoO$_2$ is an extremely liberal test case in
which the pockets are barely present in the first place. Previous LDA
calculations of the realistic structure which explicitly
included the Na, both with and without water, had already shown that LDA
predicts the survival of the pockets for
Na$_{\frac{1}{3}}$CoO$_2$\cite{Marianetti:2004B,Johannes:2004C}, even when
including full structural relaxations \cite{Marianetti:2004B}. 
Therefore, one can safely conclude that the Na potential is not the dominant
mechanism which destroys the pockets in Na$_{0.3}$CoO$_2$.

In this study, we resolve the issue of the qualitative behavior of the pockets.
We calculate the ARPES spectrum and the heat capacity at a range of different
$U$ in order to determine the best agreement with experiment. In particular, we
focus on the presence or absence of the
pockets, the linear coefficient of the heat capacity, and the average Fermi velocity
measured in ARPES. 
The presence or absence of the pockets will have important
consequences for the heat capacity given that they are the dominant contribution
to the DOS at the Fermi energy, and therefore these two issues are intimately
connected.

We perform DMFT calculations for the $t_{2g}$  bands of the
cobaltates, represented by the following Hamiltonian:  

\begin{eqnarray}\label{ham}
H=\sum_{ij\alpha\beta\sigma}t_{\alpha\beta}c^+_{i\alpha\sigma}c_{j\beta\sigma}+
\sum_{i\alpha\beta\sigma\sigma'}U_{\alpha\beta}^{\sigma \sigma'}n_{i\alpha\sigma}n_{i\beta\sigma'}
\nonumber\\
+\sum_{i\sigma}\Delta (n_{a_{1g} i \sigma} - n_{e_{g}'i \sigma})
\end{eqnarray}

where $\alpha,\beta$ are the orbital indices (ie. $a_{1g}$ and $e_{g}'$), $i,j$
are site indices, $\sigma$ is the spin index, and $\Delta$ is the crystal-field
splitting between the $a_{1g}$ and $e_g'$ orbitals.  We use the low-energy
hopping parameters $t_{\alpha\beta}$ and $\Delta$ which were fit to the LDA
$t_{2g}$ bands by Zhou et al\cite{Zhou:2005}, allowing for a direct comparison.  
Ishida et al did not publish their hopping parameters,
but they appear to be similar given the
respective bare DOS which were published\cite{Ishida:2005}. 
We assume the traditional orbital-independent double counting\cite{Kotliar:2006}
of $U(N-\frac{1}{2})$, and therefore the Hartree-Fock terms generated by DMFT
will be relevant.   Below we will show that $\Delta$ is a key parameter in
determining the fate of the $e_{g}'$ pockets  and therefore the heat capacity.
Given that LDA is only an approximate technique to generate the low energy
hopping parameters, we will systematically explore the effect of $\Delta$ on the
results.  The LDA value fit by Zhou et al is roughly $\Delta=-10 meV$, and
therefore the on-site orbital energies are nearly degenerate. Alternatively,
quantum chemistry calculations\cite{Landron:2006} yield a value of $300 meV$, so one might
anticipate $-10 meV \le \Delta \le 300 meV$.

DMFT maps the interacting lattice problem onto an impurity problem where the
non-interacting bath function is determined self-consistently
\cite{Georges:1996}.  The effective impurity problem is solved using two
different impurity solvers to ensure that the answer is robust.  The traditional
Hirsch-Fye quantum Monte-Carlo (HFQMC) method was used 
(see ref \cite{Georges:1996,Kotliar:2006} for detailed review), 
in addition to the
newly developed continuous time quantum Monte-Carlo (CTQMC) method
\cite{Werner:2006,Haule:2006}. CTQMC is generally more efficient than the HFQMC and
allows one to access significantly lower temperatures and larger U.

The $e_g'$ pockets are observed in the k-space direction corresponding to the
real-space direction which connects nearest-neighbor Co atoms.  ARPES
experiments predict the $e_g'$ pockets to be 
below the Fermi
energy\cite{Yang:2005,Qian:2006,Qian:2006B,Qian:2006C}, and this is supported by the analysis of the de Haas van Alphen
experiments
\cite{Balicas:2006}.  Given that two previous theoretical studies reach opposite
conclusions regarding the fate of the $e_g'$ pockets, we explore this question
in detail. When determining the Fermi surface from the Dyson equation, only the
bare Hamiltonian and the self-energy at zero frequency are needed
\cite{Lechermann:2005}. Within
single-site DMFT the self-energy is momentum-independent and therefore the
self-energy at zero frequency acts as a renormalization of the on-site $e_g'$
and $a_{1g}$ energy levels. Therefore, it is useful to reverse-engineer this
problem and determine what effective $e_g'$ and $a_{1g}$ levels are needed to
destroy the pockets.  This is not a-priori obvious given that the neighboring
$e_g'$-$a_{1g}$ hopping is comparable to the $a_{1g}$-$a_{1g}$ hopping.  
We find that the pockets are completely insensitive to
perturbations of the $a_{1g}$ on-site energy. 
The pockets are destroyed when the $e_g'$ on-site energy is
shifted down by roughly 70$meV$, independent of the value of the perturbation
of the $a_{1g}$ level, and therefore the
criterion for the destruction of the pockets is 
$\Sigma_{e_{g}'}-\mu<-70 meV$. This is an important observation which
indicates that a relatively small perturbation of the e$_g$' on-site energy
will destroy the pockets. DMFT calculations can now be performed to determine
when this criterion is satisfied.

\begin{figure}[htb]
\includegraphics[width=\linewidth,clip= ]{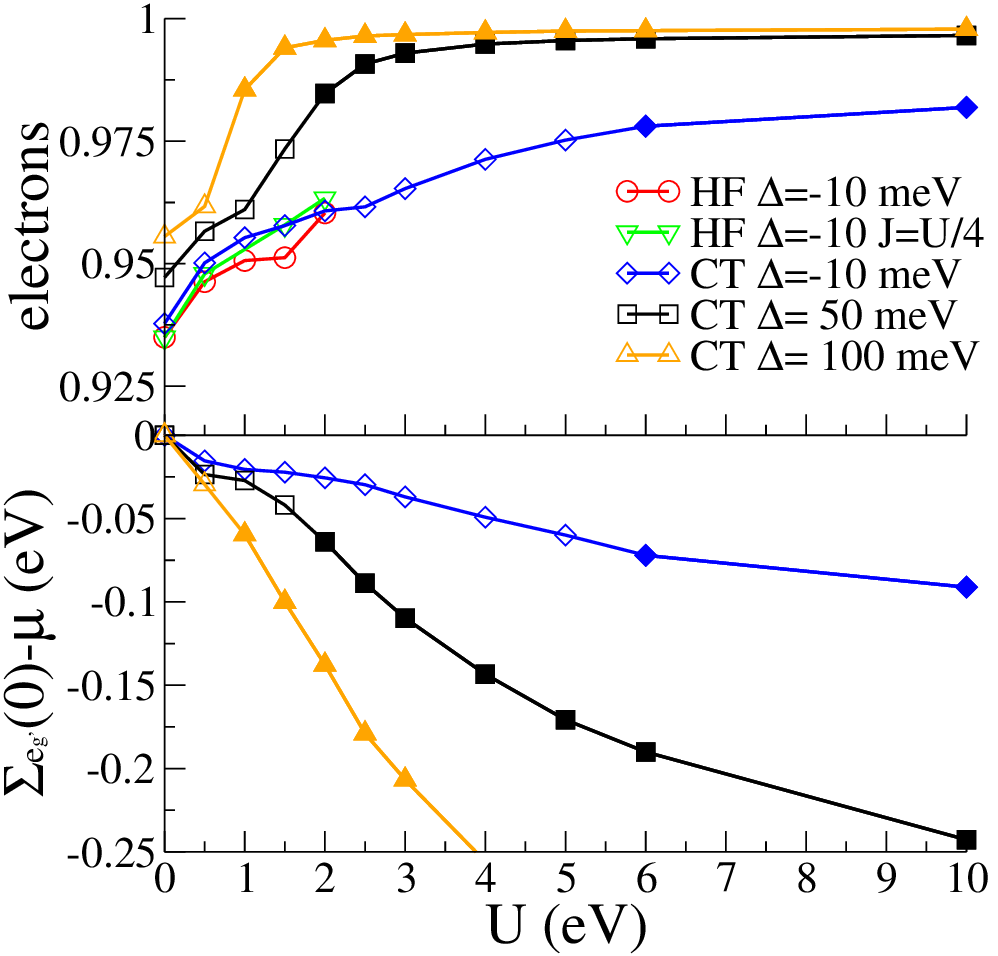}
\caption{ 
a.) $e_g'$ orbital occupation as a function of
$U$. The $a_{1g}$ orbital occupation can be found from the relation
$n_{a_{1g}}=5.3-2n_{e_g'} $  HFQMC and CTQMC calculations were performed at
$\beta=40 eV^{-1}$ and $\beta=100 eV^{-1}$, respectively. Filled points indicate that the
pockets have been destroyed.
b.)  The self-energy of the $e_g'$ orbital at zero frequency minus the
chemical potential for various values of the crystal field splitting $\Delta$
($\Delta=-10 meV$ corresponds to the LDA crystal field). 
} \label{sig}
\end{figure}

$\Sigma_{e_{g}'}-\mu$ is found to be a monotonically decreasing function of $U$,
which indicates that the pockets are \emph{diminished} with increasing
interactions (see figure \ref{sig} b).  The quantity $\Sigma_{a_{1g}}-\mu$ is a
monotonically increasing function of $U$ (not pictured).  Considering the
dynamical portion of the self-energy (ie.
$\Sigma_{dyn}(i\omega)=\Sigma(i\omega)-\Sigma(\infty)$), we find that
$\Sigma_{dyn}^{e_g'}(0)>\Sigma_{dyn}^{a_{1g}}(0)$, but this is countered by the
fact that $\Sigma^{a_{1g}}(\infty)>\Sigma^{e_{g}'}(\infty)$ and the net result
is that $\Sigma^{a_{1g}}(0)>\Sigma^{e_{g}'}(0)$.  Therefore, the destruction of
the pockets is due entirely to the static Hartree contribution to the
self-energy (ie. $\Sigma(\infty)$ ). The dynamical contribution of the
self-energy actually enhances the pockets, but ultimately the static part of the
self-energy dominates and the net effect is that increasing interactions
diminishes the pockets.  The preceding analysis is true for both the CTQMC and
the HFQMC impurity solvers. The two solvers agree completely on a qualitative
level, and are quantitatively similar for small $U$ while more significant
differences arise for larger $U$ due to discretization errors within the HF
method. We also plot the orbital occupations as a function of $U$ (see figure
\ref{sig}a), which indicates an enhancement of orbital polarization as $U$ is
increased, consistent with the destruction of the pockets. This conclusion is
found for both the CTQMC and HFQMC solvers, and this holds even when an on-site
exchange coupling (ie. $-Jn_{\alpha\sigma}n_{\beta\sigma} $ ) is included in the HFQMC
method.  Exchange had little effect on the results, in agreement with Ishida et
al \cite{Ishida:2005}.

Having established the qualitative behavior of the pockets, we continue with
more quantitative analysis.  Given that only a 70 $meV$ downward shift of the
e$_g$' on-site energy is required to destroy the pockets, we believe that it is
useful to probe the behavior of the self-energy for other values of $\Delta$ as
rationalized above.  The CTQMC calculations were repeated for $\Delta=50 meV$
and $\Delta=100 meV$. As anticipated, starting with a crystal field splitting
which diminishes the pockets acts cooperatively with interactions and causes the
system to polarize more and the pockets to be destroyed for a smaller $U$ (see
figure \ref{sig} a and b).  For $\Delta=50 meV$ the pockets are destroyed for
$U\ge2.0 eV$, while for $\Delta=100 meV$ the pockets are destroyed for $U\ge1.0 eV$.
Therefore, we conclude that the value of $U$ required to destroy the pockets
depends strongly on the value of $\Delta$.

Our findings are in qualitative agreement with Zhou et al\cite{Zhou:2005} and in qualitative disagreement
with Ishida et al\cite{Ishida:2005}. It is not clear why the results of Ishida et al are different, but it is 
likely the result of differences in the bare Hamiltonian. Above we showed that the static correlations diminish
the pockets while the dynamic correlations enhance the pockets, and the balance of these two effects
will be influenced by the bare Hamiltonian. Regardless, our following analysis of the Fermi velocity and heat capacity will
demonstrate that the destruction of the pockets is the only viable possibility to achieve agreement with experiment.

\begin{figure}[htb]
\includegraphics[width=\linewidth,clip= ]{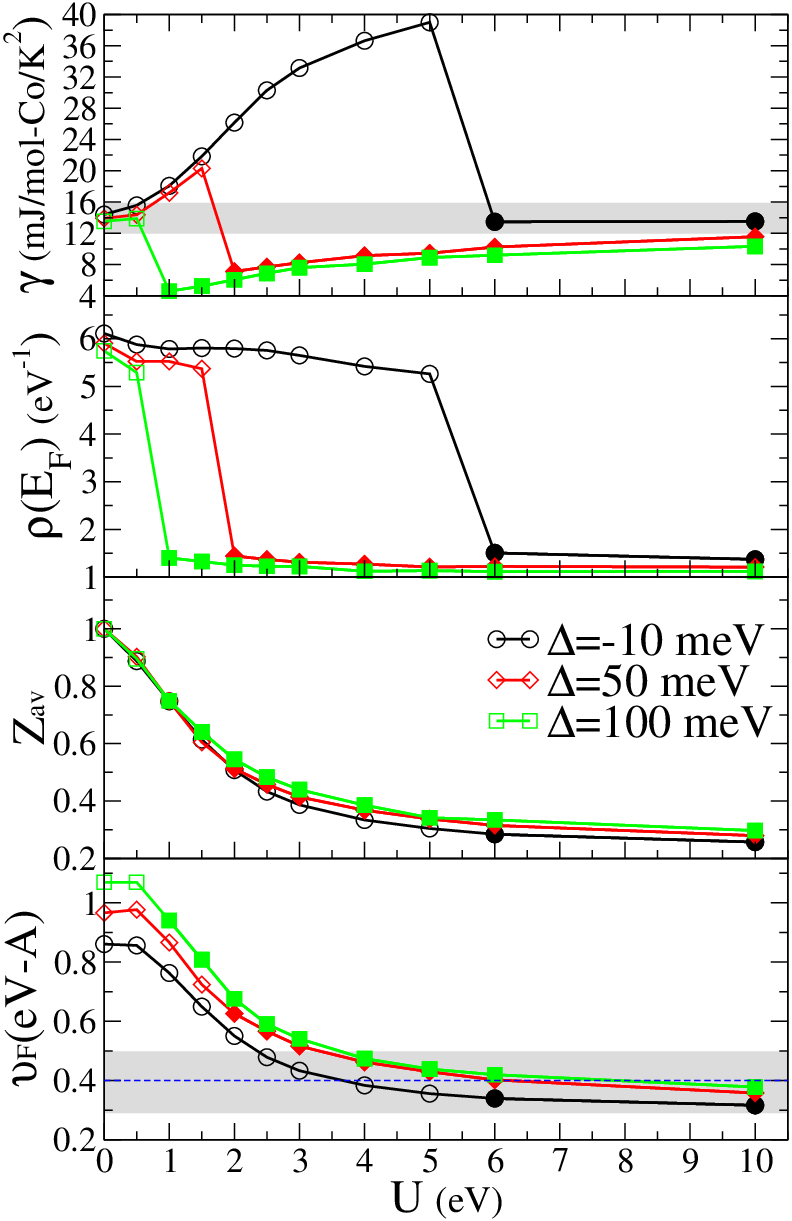}
\caption{ The heat capacity, density-of-states at the Fermi energy, the
quasiparticle weight Z, and the average of the absolute value of the Fermi velocity for the $\Gamma$-M and $\Gamma$-K directions
are shown in panels a, b, c, and d, respectively. Filled points indicate that the
pockets have been destroyed. The shaded region in panel $a$ corresponds to the
range of experimentally measured values of $\gamma$. The blue dotted line in panel d corresponds
to the experimentally measured Fermi velocity, while the grey shading corresponds to the error bar \cite{Qian:2006}.
} \label{heatcapacity}
\end{figure}

Having understood the behavior of the pockets, we now compute the linear
coefficient of the heat capacity, $\gamma$. We use Fermi-liquid theory to
calculate $\gamma$ from the DOS at the Fermi energy and the quasiparticle weight
$Z$ which is calculated in our QMC calculations.  Mathematically, we have
$\gamma=\frac{2\pi k_B^2}{3} \sum_{\alpha} \frac{\rho_{\alpha}(0)}{Z_{\alpha}}$,
where $\alpha$ corresponds to the orbital index and $\rho$ is the local spectral
function\cite{Kotliar:2006}.  The experimentally measured heat capacities for
Na$_{0.3}$CoO$_2$  are found to be in the range of 12-16
$\frac{mJ}{mol-Co-K^2}$\cite{Cao:2003,Ueland:2004,Chou:2004,Lorenz:2004,Jin:2005,Oeschler:2005}.
We begin by noting that the heat capacity is 14.27 $\frac{mJ}{mol-Co-K^2}$ for
the LDA hoppings, which is already within the bounds of experimental
measurements. This may mislead one to believe that correlations are negligible,
but a more careful examination shows otherwise.  The DOS at the Fermi energy
initially decreases weakly as $U$ increases and eventually drops in a
discontinuous fashion, which signifies the destruction of the pockets (see figure
\ref{heatcapacity}).  The quasiparticle weight also decreases as $U$ increases.
Given that the linear coefficient of the heat capacity is proportional to the
ratio $\frac{\rho(e_f)}{Z}$, the overall effect is not apriori obvious. 
The heat capacity initially increases as $U$ increases, then discontinuously
drops when the pockets are destroyed, and eventually
plateaus for large $U$. Increasing $\Delta$ causes the drop in the heat capacity
to occur at smaller values of $U$ and a overall lower value for the heat
capacity. In order to achieve agreement with experimental measurements of the 
heat capacity, one needs a relatively
large $U=6 eV$ when using the LDA $\Delta$, and even larger $U$ is needed for
larger $\Delta$. A key point is that if the pockets are retained, the $\gamma$ becomes
excessively large as $U$ increases. Given that the ARPES indicates that the bandwidth is nearly halved 
as compared to LDA, an appreciable U must be present to narrow the LDA bands and if the pockets
were still present the heat capacity would be excessive as compared to experiment.
It is reasonable to expect that the heat capacity should be
under-predicted when only considering the Hubbard model. There will likely be 
electron-phonon coupling to the local breathing mode of the octahedron, or
perhaps other modes, which will induce a narrowing of the bands and therefore an
enhancement of the heat capacity.

The experimentally measured Fermi velocity may also be calculated as a function
of $U$ (see figure \ref{heatcapacity}). 
Increasing the $U$ decreases the quasiparticle weight $Z$ and therefore decreases the Fermi velocity.
In order to achieve velocities comparable with
experiment, one needs a relatively large $U>3eV$. This is another piece of evidence, independent of the heat capacity 
measurements, which indicates that the $U$ must be relatively large. 
Once again, if the $U$ is large then the pockets must be absent in order to get acceptable agreement with the heat capacity.

\begin{figure}[htb]
\includegraphics[width=\linewidth,clip= ]{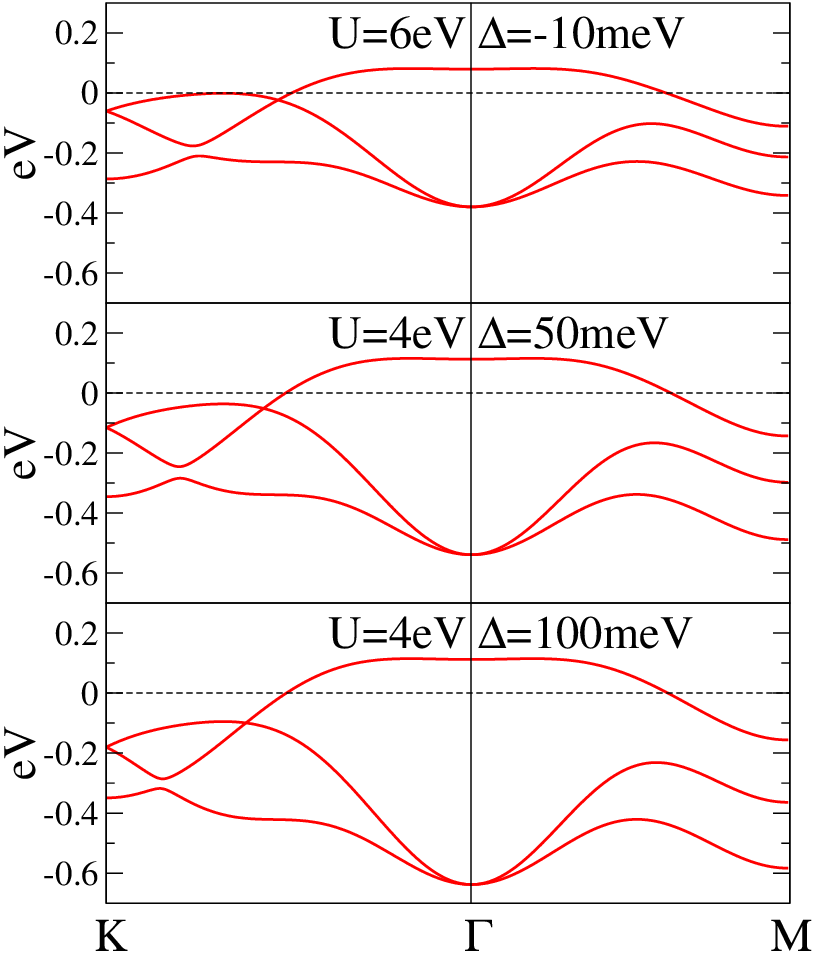}
\caption{ The  ARPES for a.) $U=6.0$ $\Delta=-10 meV$ b.) $U=4.0$ $\Delta=50 meV$ 
c.) $U=4.0$ $\Delta=100 meV$. All calculations performed at $\beta=100 eV^{-1}$. 
} \label{arpes}
\end{figure}

Both impurity solvers used in this study work on the imaginary axis, and
therefore one must perform an analytic continuation to access real frequency
quantities like the ARPES spectrum. Various approaches exist to perform the
analytic continuation, but all are approximate. We expand the self-energy to
first-order, which allows an exact analytic continuation, and use the resulting
self-energy to construct the low energy ARPES spectrum (see figure
\ref{arpes}).  The first case corresponds to the $\Delta=-10meV$ and $U=6eV$,
the minimum to destroy the pockets for this case (see figure \ref{arpes}a). As
shown, the pockets are too close to the Fermi energy, and the bands are
excessively narrow as compared to experiment. Increasing to $U=10eV$ will further
sink the pockets, but it will also to continue to narrow the bandwidth.
Alternatively, one may decrease the $U$ slightly if the $\Delta$ is increased
as the pockets are suppressed at a much lower $U$. Therefore, we examine the
ARPES for $\Delta=50meV$ and $U=4eV$, and $\Delta=100meV$ and $U=4eV$ (see
figure \ref{arpes}b and \ref{arpes}c). Decreasing $U$ and increasing $\Delta$ 
increases the bandwidth and pulls the pockets down further, putting the result in
better agreement with experiment. This analysis suggests that optimum $U$ should be 
chosen from the lower end of the range of values deduced from the analysis of the heat
capacity and the velocity, and that $\Delta$ should be slightly larger than the one
deduced from LDA.

In conclusion, we have examined the issue of the $e_g'$ pockets, the value of
the linear coefficient of the heat capacity, and the Fermi velocity.
We have demonstrated that increasing interactions destabilize the $e_g'$
pockets and pushes them beneath the Fermi energy. This is in agreement with previous calculations of
Zhou et al\cite{Zhou:2005} and in disagreement with previous calculations of
Ishida et al\cite{Ishida:2005}.  Reasonable agreement can be achieved with
both bulk heat capacity measurements and the Fermi velocity measured by ARPES when using an on-site
coulomb repulsion which is several times the LDA bandwidth (ie. $U>4 eV$). 

We acknowledge useful discussions with G. Kotliar, P.A. Lee, H. Ding, and M.Z. Hasan.
Funding was provided by NSF under grant DMR 0528969. C.A. Marianetti acknowledges funding from ICAM.

\bibliography{main}
\end{document}